\documentclass[eqsecnum,aps,prd,noshowpacs,preprintnumbers,amsmath,amssymb,nofootinbib,floatfix, superscriptaddress]{revtex4}

\usepackage[section]{placeins}
\usepackage{bm}
\usepackage{graphicx,epsf}
\usepackage{pstricks}
\usepackage{color}

\def\cs2{c_{\rm{s}}^2}

\newcommand\calp{\mathcal{P}}
\newcommand\caln{\mathcal{N}}
\newcommand\mpl{m_{\rm Pl}}

\newcommand{\fnleq}{\ensuremath{f_{\mathrm{NL}}^{\mathrm{equi}}}}
\renewcommand\({\left(}
\renewcommand\){\right)}

\newcommand\be{\begin{equation}}
\newcommand\ee{\end{equation}}
\newcommand\bea{\begin{eqnarray}}
\newcommand\eea{\end{eqnarray}}
\newcommand\eq[1]{Eq.~(\ref{#1})}

\newcommand\fig[1]{Fig.~(\ref{#1})}

\newcommand\eps{\epsilon}

\newcommand{\Vol}{\mathrm{Vol} (X_5)}

\begin{document}
\title{An update on single field models of inflation in light of WMAP7}
\author{Laila Alabidi}
\email{l.alabidi@qmul.ac.uk}
\author{Ian Huston}
\email{i.huston@qmul.ac.uk}
\affiliation{Astronomy Unit, School of Mathematical Sciences, 
Queen Mary University of London, Mile End Road, E1 4NS, UK}

\begin{abstract}
In this paper we summarise the status of single field models of inflation in light of
the WMAP 7 data release. We find little has changed since the $5$ year release, and
results are consistent with previous findings. The increase in the upper bound on 
the running of the spectral index impacts on the status of the production of Primordial
Black Holes from single field models. The lower bound on $\fnleq$ is reduced
and thus the bounds on the theoretical parameters of (UV) DBI single brane models are weakened.
In the case of multiple coincident branes the bounds are also weakened and the two, three or four
brane cases will produce a tensor-signal that could possibly be observed in the future.
\end{abstract}
\date{\today}
\maketitle

\section{Introduction}
We review the status of single-field models of inflation in light
of the latest data release from the Wilkinson Microwave Anisotropy
Probe \cite{wmap7}. We utilise the $7$ year $\rm{WMAP}$ data, combined
with the Baryon Acoustic Oscillations ($\rm{BAO}$) and measurement of the Hubble
parameter from the supernovae data, the $\rm{H}0$ set. This data set combination
is considered the best estimate for cosmological parameters at present \cite{wmap7}.

We categorise our models into `canonical' and `non-canonical' models in concordance
with Ref.~\cite{ALi}. 
Canonical models have a pressure term $P$ given as $P=X-V(\phi)$, where
$X$ is the kinetic term, $V(\phi)$ is the potential, and the inflaton ($\phi$) fluctuations
propagate at the speed of light. Non-canonical models on the other hand
have a pressure term which is non-linearly dependant on $X$ and the inflaton
fluctuations propagate at a different speed to light (see for example Refs.~\cite{GM1999, Cake}). Our canonical models
are then sub-categorised into `small' and `large' field models, where small field
models are defined as those with an inflaton variation less than the Planck scale
$\Delta\phi<\mpl$.

The observational parameters that we will be utilising in
this paper are the spectral index $n_s$, the running of the spectral index
$n_s'$, the tensor fraction $r$ and the non-gaussianity parameter
for an equilateral configuration $\fnleq$. The $\rm{WMAP}7+
\rm{H}0$ data set gives bounds on these parameters which
we list at the $2\sigma$ confidence limit,
\bea\label{obs}
0.939<n_s<0.987\,,\nonumber\\
r<0.24\,,\nonumber\\
-0.084<n_s'<0.02\,,\nonumber\\
-214<\fnleq<266\,.
\eea
We have presented the bounds on $n_s$ for
$n_s'=0$ and $r=0$ priors, the bound on $r$ is for
an $n_s'=0$ prior and the bounds on $n_s'$
are for an $r=0$ prior.

A key parameter in inflation model discrimination is the number
of $e-$folds $N$, which is the logarithmic ratio of the comoving hubble horizon
at the end of inflation to its value at the time when scales of cosmological
interest left the horizon. Liberal limits on $N$ are taken to be
\be\label{N-liberal}
10<N<110\,,
\ee
where the lower bound comes from the demand
 that nucleo-synthesis is well bounded, and the upper bound assumes 
that the universe underwent a few bouts of ``fast'' roll 
inflation \cite{Liddle:2003as}. The following bounds on $N$
are more commonly used in the literature
\be\label{N-reason}
N=54\pm7\,,
\ee
where the uncertainty comes from our ignorance of reheating.
In this paper we assume instant reheating.

In Section (\ref{can}) we briefly overview models of inflation with
canonical kinetic terms, both small and large field, and present
the results of their predictions for $n_s$, $r$ and $n_s'$ and how
they compare to the latest WMAP data release. We do not give
an in depth review of these models or their motivation, for such
reviews refer to Refs.~\cite{Lyth-Riotto,AL1,AL2,ALi,book1,book2}.
In Section (\ref{non-can}) we analyse single field Dirac-Born-Infeld
models of inflation in light of WMAP7.

\section{Canonical Models of Inflation}\label{can}
\subsection{Small Field Models}
\subsubsection{Models with negligible running}
\begin{figure}
\centering\includegraphics[ width=0.7\linewidth,totalheight=2.5in]{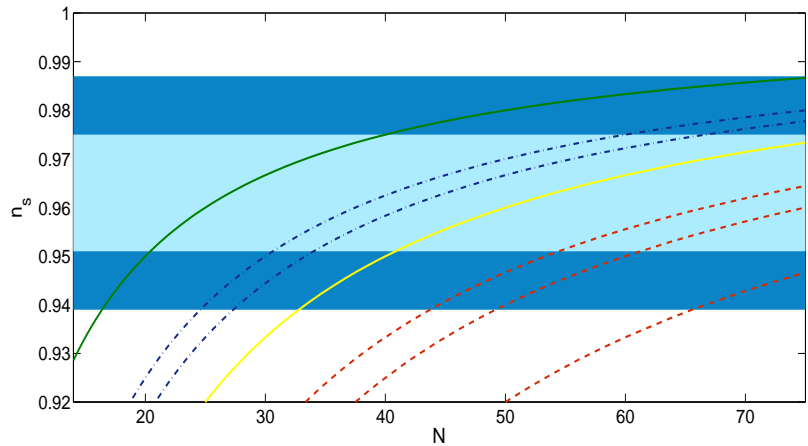}
\caption{Plot of \eq{n_small}; the spectral index versus the number of $e-$folds for fixed values
of $p$. The light blue (light grey) and dark blue (dark grey) regions correspond to the $1\sigma$
and $2\sigma$ allowed regions respectively. The dark green (uppermost) line
is the logarithmic potential (\ref{log}), below that are the blue dash-dotted $p=-3$ and $p=-4$ results
from the tree level potential (\ref{tree}), the yellow line (central line) is the
prediction for the exponential potential (\ref{exp}) and the red dashed lines are (from the bottom up)
$p=3$, $p=4$ and $p=5$ predictions of the tree level potential (\ref{tree}).}
\label{nsvN}
\end{figure}

A general form for a small field model is the tree level potential
given by
\be\label{tree}
V=V_0\left[1\pm\left(\frac{\phi}{\mu}\right)^p\right]\,,
\ee
where $\mu$ and $V_0$ are constants, $\mu\leq\mpl$ and $p$ can
be positive \cite{Linde:1981mu,Albrecht:1982wi,Linde:1984cd,Binetruy:1986ss,Banks:1995dp} 
or negative \cite{mutant}. The case $p=-4$ can also arise in certain
models of brane inflation \cite{DT}. One loop corrections in $F-$term
hybrid supersymmetry (SUSY) potentials \cite{Copeland:1994vg,Dvali:1994ms,Stewart:1994ts} also
result in a small field potential
\be\label{log}
V=V_0\left[1+\frac{g^2}{2\pi}\ln\(\frac{\phi}{Q}\)\right]\,,
\ee
which we dub the logarithmic potential. $Q$ determines
the renormalisation scale and $g<1$ is the coupling
of the super-field which defines the inflaton to the super-field
which defines the flat directions. Finally, we also analyse the exponential potential
\be\label{exp}
V=V_0\left[1-e^{-q\phi/\mpl}\right]\,,
\ee
where the value of the parameter $q$ depends on whether
\eq{exp} is derived from non-minimal inflation \cite{Salopek:1988qh} such as lifting a flat direction in SUSY via
a Kahler potential \cite{Stewart:1994ts} or from non-Einstein gravity \cite{Starobinsky:1980te,Lyth-Riotto}. This
potential also arises from assuming a variable Planck mass (see for example \cite{Spokoiny:1984bd, Bardeen:1987zb})
and from Higgs inflation (see for example \cite{Bezrukov:2007ep}).Though
not technically a small field model, since $\phi$ can be greater 
than $\mpl$ in the region where $V_0$ dominates, we group
this potential in this category because it predicts a small $r$ and
shares the same form for the spectral index as the other small-field models,
\be\label{n_small}
1-n_s=\frac{2}{N}\(\frac{p-1}{p-2}\)\,.
\ee
\eq{log} corresponds to $p\to0$ and \eq{exp} corresponds
to $p\to\infty$. We plot the results for \eq{n_small} in \fig{nsvN},
for the range of e-folds $14<N<75$. We find that the prediction
for the model with
$p=3$ does not intersect with the $1\sigma$ region for this range;
we would require more than $67$ e-folds of inflation for it to intersect
with the $2\sigma$ region and more than $80$ for it to intersect with
the $1\sigma$ region. If we were to limit ourselves to the standard
range of $e-$folds (\ref{N-reason}) then the positive powers of $p$ will be further constrained
by the data, as summarised in Table~(\ref{table}).
\begin{table}[ht]
\centering
\begin{tabular}{|c|c|c|c|}
\hline
&Outside the&Outside the\\
&$1\sigma$ region&$2\sigma$ region\\
\hline
$N=47$&$p<14.4$&$p<4.6$\\
\hline
$N=54$&$p<6.1$&$p<3.7$\\
\hline
$N=61$&$p<4.8$&$p<3.2$\\
\hline
\end{tabular}
\caption{The exclusion limits for positive values of $p$ for particular 
values of $N$, 
based on the combined ${\rm WMAP7}+{\rm BAO}+{\rm H_0}$ data on the 
spectral index. All values of 
$p<0$ are included at the $1$ and $2\sigma$ levels for this range.}
\label{table}
\end{table}

\subsubsection{Models which predict PBHs}
Black holes could have been produced at the end of inflation with an abundance that may be detected,
if the spectral amplitude at the corresponding scales is $\calp_\zeta\sim10^{-3}$ \cite{Carr-Hawking, Khlopov, Zaballa, Kohri-Lyth}.
It has been shown \cite{Kohri-Lyth} that this requires that the running of the spectral index is positive 
$n_s'>0$  and that $\epsilon(\phi_{end})<\eps(\phi_*)$,
which is predicted by a hilltop-type model of inflation \cite{AK} and
the running mass model \cite{rmm1,rmm2,rmm3}. The upper bound on the spectral
amplitude at the end of inflation is given as $\calp_\zeta(N=0)<0.03$ \cite{josan,carr-kohri},
and the upper bound on $n_s'$ has gone up to $0.02$. Defining
$\mathcal{B}=\eps(\phi_e)/\eps(\phi_*)$
the condition for PBH formation is then
\be
-8<\log\mathcal{B}<-6\,.
\ee

The Hilltop-type inflation model is given by the phenomenological
potential
\be\label{hill}
V=V_0\(1+\eta_p\(\frac{\phi}{\mpl}\)^p-\eta_q\(\frac{\phi}{\mpl}\)^q\)\,,
\ee
where $0<p<q$, and the Running mass model is given by
\bea\label{rmm}
V&=&V_0\left[1-\frac{\mu_0^2+A_0}{2}\(\frac{\phi}{\mpl}\)^2\right.\nonumber\\
&&\left.+\frac{A_0}{2(1+\alpha\ln(\phi/\mpl))^2}\(\frac{\phi}{\mpl}\)^2\right]\,,
\eea
where $\mu_0^2$ is the mass of the
inflaton squared, $A_0$ is the gaugino mass squared in units of $\mpl$, and $\alpha$ is related
to the gauge coupling. 

\begin{figure}
\centering\includegraphics[ width=0.7\linewidth,totalheight=2.5in]{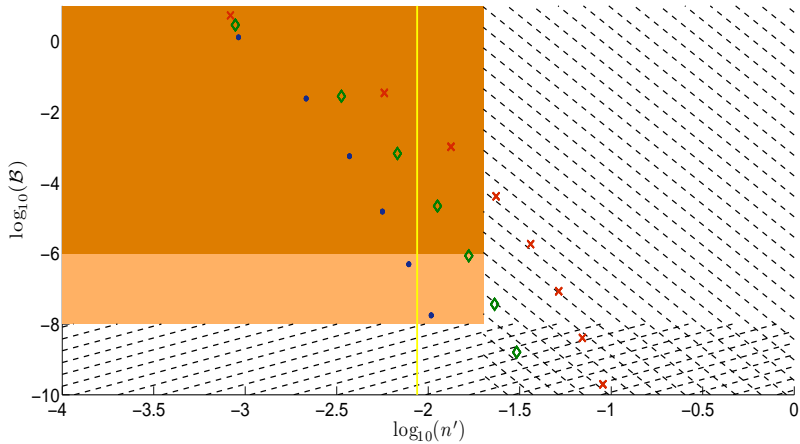}
\caption{Plot of $\log(\mathcal{B})$ vs. $\log(n_s')$ for the hilltop model (\ref{hill}) with 
$N=68$ and a range of inflaton couplings. The  blue dots represent $\{p,q\}=\{2,2.5\}$,
the green diamonds $\{p,q\}=\{2,3\}$ and the red crosses are $\{p,q\}=\{2,4\}$. The vertical yellow
line corresponds to the WMAP5 upper bound on $\log(n_s')$. The hatched region is excluded,
representing
$\log(\mathcal{B})<-8$ and $\log(n_s')>-2$. The region $\log(\mathcal{B})>-6$ does not lead to the 
formation of PBHs with an abundance that can be detected, and is represented by the 
tan color (dark region) in the figure. PBHs can form in the region 
$-8<\log(\mathcal{B})<-6$ with an abundance that may be detected and without 
violating astrophysical or cosmological bounds; it is represented by the light orange region.}
\label{pbh_hill}
\end{figure}

We have analysed \eq{hill} for a range of inflaton couplings with fixed $N=68$ and 
$n_s=0.924$, and we plot the results in \fig{pbh_hill}. We find that both
$\{p,q\}=\{2,2.5\}$ and $\{p,q\}=\{2,3\}$ can lead to the formation of PBHs without
violating astrophysical constraints, if $\eta_p\ll1$. We also find that $\{p,q\}=\{2,4\}$
would lead to the formation of PBHs for $N=100$. As for the running mass model
we find that the extension of the upper bound on $n_s'$ does not change the
conclusions found in Ref.~\cite{AK}. This is because the conservative $N$ bound,
which was used to place bounds on $\mu_0^2$ and $A_0$, coincides with a running
$n_s'<0.01$, imposing $n_s'=0.02$ means that scales of cosmological interest would have had to
exit the horizon $N\sim20$ $e-$folds prior to the formation of PBHs, for $\alpha=0.01$.

\subsection{Large field models}

Since inflation generically predicts a primordial gravitational wave background 
a detection of this signal would provide strong evidence in favour of inflation \cite{Starobinsky:1979ty}.
The signature of gravitational waves is parametrised by the tensor fraction $r$, the analytical
form for which was first derived by Ref.~\cite{Starobinsky:1985ww} in which the fact that
models of the form $V(\phi)\propto\phi^n$ lead to a large $r$ was highlighted. This result was
extended by Ref.~\cite{lyth-bound}, where the author found that large field models
necessarily lead to a significant tensor fraction, and is why such models are 
of particular interest.
%Therefore, large field models are of particular interest since they necessarily lead to a significant
%tensor fraction \cite{lyth-bound}. 
The basic class of large field models
is the monomial potential $V\propto\phi^\beta$, where $\beta$ can be
a positive (chaotic) \cite{Linde:1983gd, McAllister} or negative (intermediate) \cite{Barrow:1990vx} integer, 
or a positive fraction (monodromy) \cite{Silverstein:2008sg}. The dependence of 
the tensor fraction on the spectral index is then
\bea
1-n_s=\frac{2+\beta}{2N}\,,&&r=\frac{8}{N}\(N(n_s-1)-1\)\,,\label{chaotic}\\
r&=&\frac{8|\beta|}{|\beta|-2}\(1-n_s\)\,,\label{inter}
\eea
where \eq{chaotic} is for $\beta>0$ and \eq{inter} is for $\beta<0$.
A realisation of the chaotic, $\beta=2$, model arises in Natural-Inflation \cite{Freese:1990rb,Adams:1992bn}, where
the inflaton is represented by a pseudo-Nambu Goldstone boson, with a sinusoidal
potential
\be\label{nat}
V=\frac{V_0}{2}\(1+\cos\(\frac{\phi}{\mu}\)\)\,.
\ee

\begin{figure}
\centering\includegraphics[ width=0.7\linewidth, totalheight=2.5in]{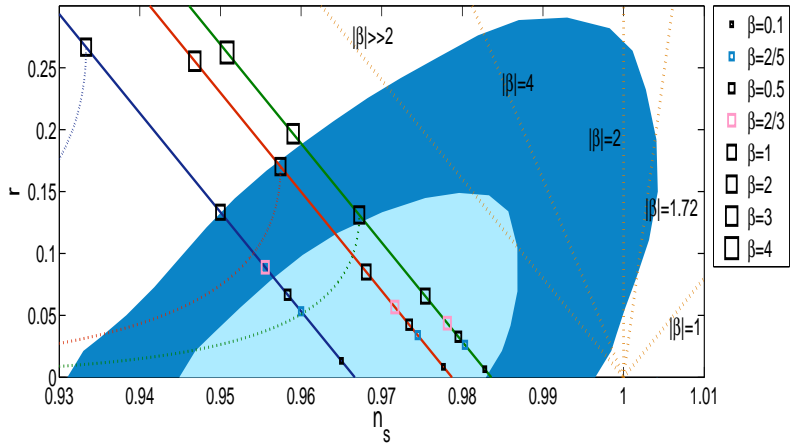}
\caption{Plot of the tensor fraction versus the spectral index for the monomial and natural inflation
models. The light blue (light grey) and dark blue (dark grey) regions correspond to the $1\sigma$ and $2\sigma$
regions respectively. The solid straight lines from left to right correspond to $N=47$ (dark blue), $N=54$
(dark red) and $N=61$ (dark green), the curved lines which intersect these lines represent the predictions
for natural inflation for the corresponding number of $e-$folds. The blue and pink squares correspond to
$\beta=2/5$ and $\beta=2/3$ respectively, and are the predictions from the Monodromy model \cite{Silverstein:2008sg}.
The dashed lines are the predictions for the intermediate models, $\beta<0$. The data used is the $\rm{WMAP}7 + \rm{H}0$ 
set applied to the $\Lambda\rm{CDM}+\rm{tensor}$ model with a zero running prior. The contours were 
generated using the Matlab scripts provided by the Cosmomc package.}
\label{rvn}
\end{figure}

We plot the tensor fraction versus the spectral index for the large field
potentials in \fig{rvn}. We find that the intermediate model with $|\beta|\gg2$
is now allowed at $1\sigma$.

\section{Non-canonical models of inflation}\label{non-can}
\subsection{Single brane DBI models}
Non-canonical models are defined as having a pressure term which
is a non-linear function of the kinetic term. In these models inflation can occur
for a non-flat potential or a non-slowly rolling inflaton. The former scenario
results in a near exponential expansion of the universe if the velocity
of the rolling inflaton is restricted, as is the case in the Dirac-Born-Infeld model.
In this type IIB string theory inspired model, $D(3+2n_d)$ branes propagate in the warped throats
of a Calabi-Yau Manifold \cite{DT,HenryTye:2006uv} in what is termed the Ultra Violet
(UV) case, where $n_d$ parametrises the dimensionality of the brane. The inflaton in this case
is the radial position of the brane with respect to the tip of the throat, and the motion
of the brane, and hence the rate of $\phi$ variation is restricted by the warped geometry  \cite{Giddings:2001yu,Silverstein:2003hf,Alishahiha:2004eh}.
In the setup we are considering, the base of the throat is an Einstein-Sasaki manifold
$\rm{X}_5$, while the throat is an $\rm{AdS}_5\times{}X_5$ manifold with a Klebanov-Strassler
background geometry \cite{KS}.  The pressure term is given by \cite{KOB}
\be\label{dbi}
P_{2n_d+3}=-T_{2n_d+3}h^4\left(1-\frac{2X}{T_{2n_d+3}h^4}\)^{1/2}-V(\phi)+T_{2n_d+3}h^4\,,
\ee
where $n_d=(0,1,2,3)$, $T_3$ is the tension of a $D3$ brane, $h$ is the throat
warp factor and $T_3h^4$ is the tension of the brane. 
One can then derive a modified consistency relation \cite{ALi}
\bea\label{consist}
Q&=&r\sqrt{-f_{NL}\calp_\zeta}-\frac{4}{\sqrt{3}}(1-n_s)\sqrt{\calp_\zeta}\nonumber\\
&=&\frac{1}{\sqrt{3}}\left[\frac{2^{15+n}}{\pi^3v_{2n}}\frac{1}{g_s^{n/2}}\(\frac{\rm{Vol}(\rm{X_5})
}{\caln}\)^{(2+n)/2}\right]^{1/4}\,,
\eea
where $v_{2n}$ is the volume of the wrapped throat, $g_s$ is the string
coupling. $\rm{Vol}(\rm{X}_5)$ is the base volume and $\caln$ is related to
the geometry of the throat.

Since $n_s<1$, then from \eq{consist} we get an upper bound on the string-theoretic
parameters
\be
Q<r\sqrt{-f_{NL}\calp_\zeta^{1/2}}<0.1\,,
\ee
where we used $f_{NL}=-214$, $\calp_\zeta=2.5\times10^{-9}$ and $r=0.25$
to get the upper bound on $Q$. This is a greater value, and hence a weaker bound than that
found in Ref.~\cite{ALi} using the $\rm{WMAP}5$ data.

For the wrapped $D5$ and $D7$ branes this gives upper bounds on
the ratio $\rm{Vol}(\rm{X}_5)/\caln$ that are independent of the inflationary
potential
\bea\label{dbi_bounds}
\left.\frac{\rm{Vol}(\rm{X}_5)}{\caln}\right|_{D5}&<&7.4\times10^{-6}g_s^{1/3}\,,\nonumber\\
\left.\frac{\rm{Vol}(\rm{X}_5)}{\caln}\right|_{D7}&<&1.5\times10^{-4}g_s^{1/2}\,.
\eea
which again are weaker upper bounds than those found in Ref.~\cite{ALi}, and
is a result of a weaker bound on $\fnleq$.
The small values in \eq{dbi_bounds} would not be  cause for concern if they
were consistent with theoretical expectations. However, theory suggests that 
$\caln<75,852$ \cite{KLEMM} and $\rm{Vol}(\rm{X}_5)\sim{}O(\pi^3)$ \cite{KS}.
If we were to satisfy the bound on $\caln$, this would
imply for the $D5$ brane that $\rm{Vol}(\rm{X}_5)<0.5$ and $\rm{Vol}(\rm{X}_5)<\pi^3/3$ for the $D7$
brane. On the other hand satisfying the bound on $\rm{Vol}(\rm{X}_5)$ requires
$\caln>4,200,000$ for the $D5$ brane and $\caln>206,700$ for the $D7$ brane.

%
% % % % % % % % % % % % % % % % % % 
\subsection{Multiple Branes}
Models with multiple branes moving in a throat have also been postulated. When $n$ branes are
separated by equal distances and follow the same trajectory, the resultant theory is equivalent to
$n$ copies of the action for a single brane. More generally however multiple branes are expected to
be separated over a range of scales with some being coincident. In Ref.~\cite{thomasward} a
model of multiple D3-branes was proposed in which the branes are coincident and the action exhibits
a non-Abelian structure. In the relativistic limit of small sound speed the action for a small
finite number of coincident branes takes the form \cite{HLTW}
\be
P = 2T_3 \left\{ h^4\sqrt{1 + (n-1)^2Y}\left(1 - \frac{\dot{\phi}^2}{T_3 h^4} \right) \right\} 
- nT_3(V-h^4)\,,
\ee
where $Y$ is defined as
\be
Y \equiv \frac{8\pi}{(n-1)^4}\frac{g_s}{T_3}\left(\frac{\phi}{h}\right)^4\,.
\ee

The equilateral non-gaussianity parameter can be written in terms of derivatives of the action
with respect to $\dot{\phi}$. We will consider only that part of the branes' motion in the throat
during which the scales leaving the horizon are those for which observational limits are placed on
the tensor signal. For WMAP7 the number of e-folds this corresponds to, $\Delta N_*$, is around
four. By using the Lyth bound \cite{lyth-bound} and relating the scalar power spectrum and
derivatives of the action, an upper bound on the tensor to scalar ratio during observable inflation
can be derived \cite{HLTW}:
\begin{equation}
\label{eq:fullbound}
r_* < \frac{1100}{(\Delta N_*)^6} \frac{[1+(n-1)^2 Y]}{\Vol} \calp_\zeta (\fnleq)^2 \,.
\end{equation}
This bound is valid whenever the change in $\phi$ over observable scales is smaller than the
magnitude of $\phi$ when the observable section begins, or $\phi_* > \Delta\phi_*$. This is true
for all UV models, with branes propagating down the throat, but must be assumed in the IR case when
branes propagate away from the tip towards the bulk.

If as before the throat is an $\mathrm{AdS_5}\times\mathrm{X_5}$ manifold then $Y$ takes a constant
value 
\begin{equation}
 Y_\mathrm{AdS} \equiv \frac{4\pi^2 g_s \caln}{(n-1)^4 \Vol}\,.
\end{equation}
Substituting this into Eq.~\ref{eq:fullbound}, using the values $\Vol\simeq \pi^3$ and
$g_s\simeq10^{-2}$, and saturating the $\caln<75852$ bound gives
\begin{equation}
 r_* < 2.8 \times 10^{-8} \left( \frac{\fnleq}{n-1} \right)^2 \,.
\end{equation}

\begin{figure}
\centering\includegraphics[totalheight=2.5in]{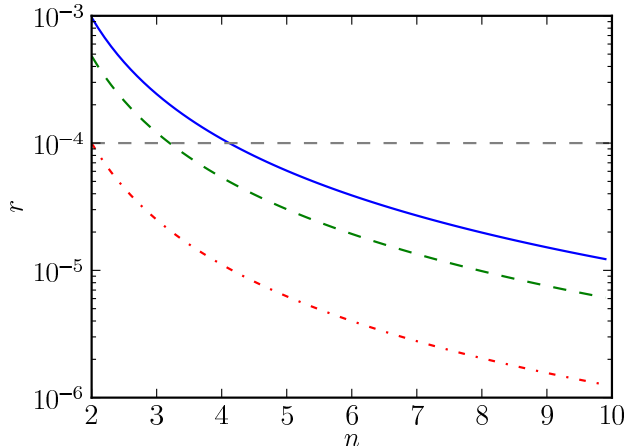}
\caption{The tensor to scalar ratio is bounded from above by a function of $\fnleq$ and the
number of branes. The upper blue line plots this relationship when the WMAP7 bound
$\fnleq>-214$ is saturated. The green dotted line is the equivalent for the WMAP5 value of the
bound, $\fnleq>-151$. The red dash-dotted line plots the relationship if $\fnleq\simeq
70$, in which case an observable tensor signal of $r>10^{-4}$ (shown by the grey dashed line) is
not possible for any number of coincident branes.}
\label{nbranes_fnl}
\end{figure}

In Fig.~\ref{nbranes_fnl} the upper bound on $r$ is plotted against the number of coincident
branes for the 2-$\sigma$ limit on $\fnleq$ from WMAP5 and 7. As the bound on $\fnleq$ has widened
considerably in WMAP7, the possible tensor-scalar ratio has increased. Setting an optimistic
observable limit of $r>10^{-4}$ limits the number of coincident branes which would produce an
observable signal. For WMAP5 an observable signal is limited to the two or three brane cases. With
the WMAP7 limits the four brane case is also included. It is worth remarking that should the
observational bounds on $\fnleq$ be reduced to $|\fnleq|<70$ then the multi-coincident brane model
would not produce an observable tensor signal for any number of branes.

\section{Summary}
In this paper we have analysed single field models of inflation with both
canonical and non-canonical kinetic terms and find that the results
are consistent with those of previous analyses using WMAP 5 data \cite{ALi,Finelli:2009bs}. 
In Ref.\cite{Finelli:2009bs} the authors use WMAP5 data 
in combination with ACBAR, QUAD, BICEP and SdSS LRG7 data and analyse
models both with a zero and non-zero running. They conclude that there is some statistical preference
of running models over power law models. In contrast, in this paper we have used the WMAP7 data combined
with the BAO and Supernovae data sets and have not analysed models with a non-zero spectral running. 
We have updated the bounds
on the relevant model parameters, and show that for the canonical
models of small-field inflation, the power of the self-coupling in the tree-level
potential with $p>0$ is more tightly constrained with the $p=3$ model requiring more than $67$
$e-$folds of inflation to satisfy the data at $2\sigma$, and the logarithmic potential
requiring less than $40$ $e-$folds of inflation to satisfy the data at $1\sigma$.
Since the upper bound on the running of $n_s$ has increased, this has
expanded the parameter space which allows for the production of PBHs
within astrophysical bounds. The hilltop-type model of inflation
now leads to the formation of PBHs for $\{p,q\}=\{2,3\}$ for $N=68$ $e-$folds
of inflation. We also find that imposing $N>20$ $e-$folds on the running mass model 
excludes the parameter space that leads to $n_s'>n_s'|_{WMAP5}$, and as such 
we find no change from the conclusions of Ref.~\cite{AK}. We still find that the formation
of PBHs is strongly dependent on the allowed upper bound on $N$, consistent with
Refs.~\cite{PE,AK}. However, the fact
that PBHs may form after less than $20$ $e-$folds of inflation in this model, raises
the question of whether this model leads to the overclosure of the universe, and may
be an avenue for further research.
The monomial potential
with positive power is still consistent with data at $1$ and $2\sigma$ for even 
a conservative range of $N$, and we find that we can now rule in the intermediate
model with a power much greater than $2$ at the $1\sigma$ level.

We find that the limits on the theoretical parameters of DBI models, with a single brane 
falling into a warped throat towards the tip, are weakened, however theoretical
expectation is still at odds with observational bounds. In order to `marry' theory
to observation we would need to motivate a smaller base volume, a larger
Euler number or observe a more negative $\fnleq$. Ref.~\cite{Huang:2010up} included
the bounds on \emph{local} $f_{\rm{NL}}$ in their analysis of the single field DBI model,
and as a result exclude it from the $1\sigma$ regime.
We also find that the bounds on $\fnleq$ and $r$ now allow for up to $4$
branes in multi-brane DBI inflation.

\acknowledgements{We would like to thank Karim Malik for useful discussions
and comments. We acknowledge use 
of the Cosmomc Matlab scripts. LA is supported by the 
Science and Technologies Facilities Council (STFC) under Grant 
PP/E001440/1. IH is supported by the STFC under Grant ST/G002150/1. }
\bibliographystyle{apsrev}
\bibliography{wmap7}

\end{document}